\documentclass[conference]{IEEEtran}
\IEEEoverridecommandlockouts
\usepackage{cite}
\usepackage{amsmath,amssymb,amsfonts}
\usepackage{algorithmic}
\usepackage{graphicx}
\usepackage{subcaption}
\usepackage{textcomp}
\usepackage{xcolor}
\usepackage{acronym}

\DeclareMathAlphabet\mathbfcal{OMS}{cmsy}{b}{n}

\def\BibTeX{{\rm B\kern-.05em{\sc i\kern-.025em b}\kern-.08em
    T\kern-.1667em\lower.7ex\hbox{E}\kern-.125emX}}

\acrodef{AI}[AI]{Artificial Intelligence}
\acrodef{IoT}[IoT]{Internet of Things}
\acrodef{DNN}[DNN]{Deep Neural Network}
\acrodef{M2M}[M2M]{machine-to-machine}
\acrodef{MI}[MI]{mutual information}

\usepackage{fancyhdr}
\fancypagestyle{firstpage}{
  \fancyhf{}
  \fancyhead[C]{\small To be presented at the \textit{Picture Coding Symposium (PCS)}, San Jose, CA, USA, December 7 - 9, 2022.}
}

\begin{document}

\title{Privacy-Preserving Feature Coding for Machines
}

\author{\IEEEauthorblockN{Bardia Azizian and Ivan V. Baji\'{c}}
\IEEEauthorblockA{School of Engineering Science, Simon Fraser University, Burnaby, BC, Canada }
}


\maketitle

\begin{abstract}
Automated machine vision pipelines do not need the exact visual content 
to perform their 
tasks. Therefore, there is a potential to remove 
private information from the data 
without significantly affecting the machine vision accuracy. 
We present a novel method to create a privacy-preserving latent representation of an image that could be used by a downstream machine vision model. 
This latent representation is constructed using adversarial training to prevent accurate reconstruction of the input while preserving the task accuracy. Specifically, we split a \ac{DNN} model 
and insert an autoencoder whose purpose is to both reduce the dimensionality as well as remove  information relevant to input reconstruction while minimizing the impact on task accuracy. 
Our results show that input reconstruction ability can be reduced by about 0.8 dB at the equivalent task accuracy, with degradation concentrated near the edges, which is important for privacy. At the same time, 30\% bit savings are achieved compared to coding the features directly. 
\end{abstract}


\thispagestyle{firstpage}

\section{Introduction}
The growth of \ac{AI} applications such as \ac{IoT}, visual surveillance, autonomous driving, industrial machine vision, etc., has resulted in a proliferation of ``intelligent'' edge devices, sensors, and their associated infrastructure. 
These 
devices need to communicate with each other and the cloud-based services 
to accomplish specific tasks. According to Cisco's annual Internet report~\cite{cisco_report}, most globally connected devices will be allocated to \ac{M2M} connections by 2023. Accordingly, significant research effort is being devoted to the development of efficient data coding for machines. Related standardization activities such as JPEG AI~\cite{JPEG-AI} and MPEG-VCM \cite{MPEG-VCM} have also been initiated.

With the massive amounts of data being passed around among edge devices and the cloud, privacy concerns naturally arise. What happens if a malicious third party gains access to this data? Depending on the particular scenario and application, there are a number of privacy and security concerns, exemplified through various attacks~\cite{privacy_survey,mireshghallah_privacy_2020}. Our particular focus in this paper is on the \textit{model inversion attack}~\cite{mia_2019}, where the attacker attempts to recover the original image from the intercepted features. If such an attack is successful, the attacker obtains the original image containing private information, which they may then utilize for malicious purposes. 
Cryptography methods~\cite{cryptography_cst2016} provide one possible solution to protect the data, although they have their own risks and challenges. But, in the context of data coding for machines, there is an opportunity to reduce or remove private information from data while retaining task-relevant information. This is because machine vision  generally needs higher-level information, rather than details of each pixel, in order  to perform a given task. For instance, the details of a vehicle's license plate or a person's face are not necessary if the \ac{DNN} model is only supposed to detect cars and pedestrians on the street.

In this paper, we propose a feature coding method for machines that allows object detection at near-default accuracy (i.e., close to the model's accuracy without compression) at high bitrates, while being resistant to model inversion attacks. 
To do so, we train two networks simultaneously in an adversarial manner~\cite{GAN}. One is part of an object detection pipeline, and the other is an adversary trying to perform input reconstruction from encoded features. The loss functions used in training are designed to encourage high object detection accuracy and low input reconstruction performance, especially near the edges because those details often reveal private information.

The paper is organized as follows. Related work is discussed in Section~\ref{sec:rel_works}. A comprehensive explanation of the proposed method is given in Section~\ref{sec:method}. Section~\ref{sec:exp_res} presents our experimental results, followed by conclusions in Section~\ref{sec:conclusion}.

\begin{figure*}[ht!]
    \centering
    \includegraphics[width=0.85\textwidth]{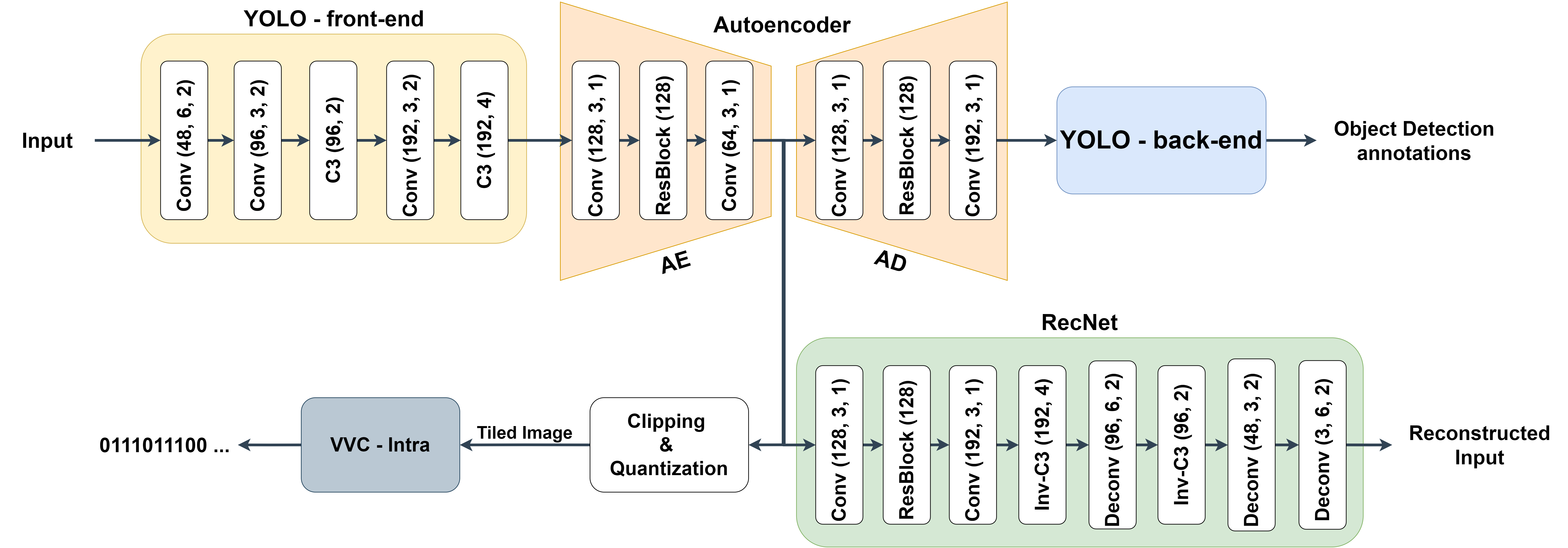}
    \caption{The overall block diagram of the proposed method. $Conv(n,k,s)$ is a 2D Convolution layer followed by a Batch normalization layer and a $SiLU()$ activation, with $n$ being the number of output channels, a kernel size of $k\times k$, and stride=$s$. $Deconv(n,k,s)$ is the same as $Conv(n,k,s)$ except with a Convolution Transpose layer. Note that the $Conv$ layers in the autoencoder do not have batch normalization, and the last $Conv$ in AE does not have the $SiLU()$ activation either. The structure of other blocks is shown in Fig.~\ref{fig:sub_blocks}}.
    \label{fig:block_diagram}
    \vspace{-0.4cm}
\end{figure*}

\section{Related Work}
\label{sec:rel_works}
For decades, image codecs have been designed to improve compression efficiency by decreasing the number of bits representing the input while maintaining the fidelity between the original and reconstructed input. This principle works well for human vision, especially when the fidelity is measured using  perceptual metrics~\cite{Lin_Kuo_JVCIR_2011}. In recent years, DNN models have entered the area of image coding~\cite{balle_hyperprior, balle_autoregressive, Cheng_2020_CVPR}, showing steady progress in comparison with traditional codecs. 

Increasingly, however, much of the visual content is only ``seen'' by machines in applications such as autonomous driving and navigation, traffic monitoring, etc. In such cases, it may be advantageous to code task-relevant features derived from images instead of images themselves. The utility of feature coding has been recognized even prior to the current wave of interest in \acp{DNN}, through MPEG standards on Compact Descriptors for Visual Search (CDVS)~\cite{cdvs_std} and Compact Descriptors for Visual Analysis (CDVA)~\cite{cdva_std}. More recently, feature compression has been studied in the context of collaborative intelligence~\cite{CI_ICASSP_2021}, where a \ac{DNN} is distributed between an edge device and the cloud, and features need to be uploaded from the edge to the cloud to complete the inference. This scenario is also assumed in the present paper.


Like images, features derived from intermediate \ac{DNN} layers can be encoded either using traditional or \ac{DNN}-based codecs. Early work on this topic~\cite{Collab1_hyomin, bottlenet, compressible_loss} preferred traditional codecs; such approaches are still appealing due to traditional codecs' computational simplicity relative to \ac{DNN}-based codecs, and their wide availability in existing cameras and devices. In order to use a conventional codec for feature coding, the feature tensor usually needs to be tiled into an image, scaled, and pre-quantized~\cite{Collab1_hyomin}. 
The authors of~\cite{bottlenet} additionally reduced the dimensionality of the latent space in terms of both the number of channels and spatial resolution using an autoencoder prior to coding the bottleneck tensors via JPEG. Such dimensionality reduction usually helps compression efficiency. 


More recent schemes~\cite{End_learned_machines, YOLOv5_based_coding} employ DNN-based coding tools, especially advanced entropy models, to code features. An advantage of such schemes is their ability to be trained end-to-end,  with a loss function that combines a differentiable rate estimate and task accuracy. Specifically,~\cite{End_learned_machines,YOLOv5_based_coding} uses the scale hyperprior entropy model from~\cite{balle_hyperprior} to obtain the rate estimate. While the ability to train end-to-end affords additional flexibility to these methods, the downside is increased complexity and the fact that their coding engines are not widely available in existing devices.  

None of the above-mentioned feature coding methods has considered privacy issues related to coded features. In fact, research on privacy protection in the context of feature coding for machines is fairly scarce. 
One of the rare works on this topic is~\cite{Privacy_Saeed}, where an information-theoretic privacy approach called \textit{privacy fan} is proposed. Here, features from a \ac{DNN} are scored according to the \ac{MI}~\cite{Cover_Thomas_2006} relative to the private and non-private tasks. Features with less private and more non-private information are only lightly compressed, whereas those that carry more private information are compressed more heavily to protect privacy. 

At its core, privacy fan is a feature selection method based on \ac{MI} between features and private/non-private tasks, which is difficult to estimate in high-dimensional spaces~\cite{Kraskov_2004}. The approach presented in this paper avoids this challenge by using an autoencoder. For this reason, it is also more flexible than the privacy fan because it enables not just feature selection, but also feature modification. Details are presented next.


\section{Proposed Method}
\label{sec:method}
The main goal of this research is to develop a framework for feature coding for machines, which has improved resistance to model inversion attacks. Our pipeline is shown in Fig.~\ref{fig:block_diagram}. We chose object detection by YOLOv5~\cite{glenn_jocher_2022_6222936} as the machine task, but the framework is applicable to other \ac{DNN} models and tasks as well. 
Several models with varying complexity and accuracy are available in the YOLOv5 Github repository.\footnote{https://github.com/ultralytics/yolov5} We chose YOLOv5m and set the ``\textit{image-size}'' parameter to 512. 

\begin{figure}[t!]
    \centering
    \includegraphics[width=0.8\columnwidth]{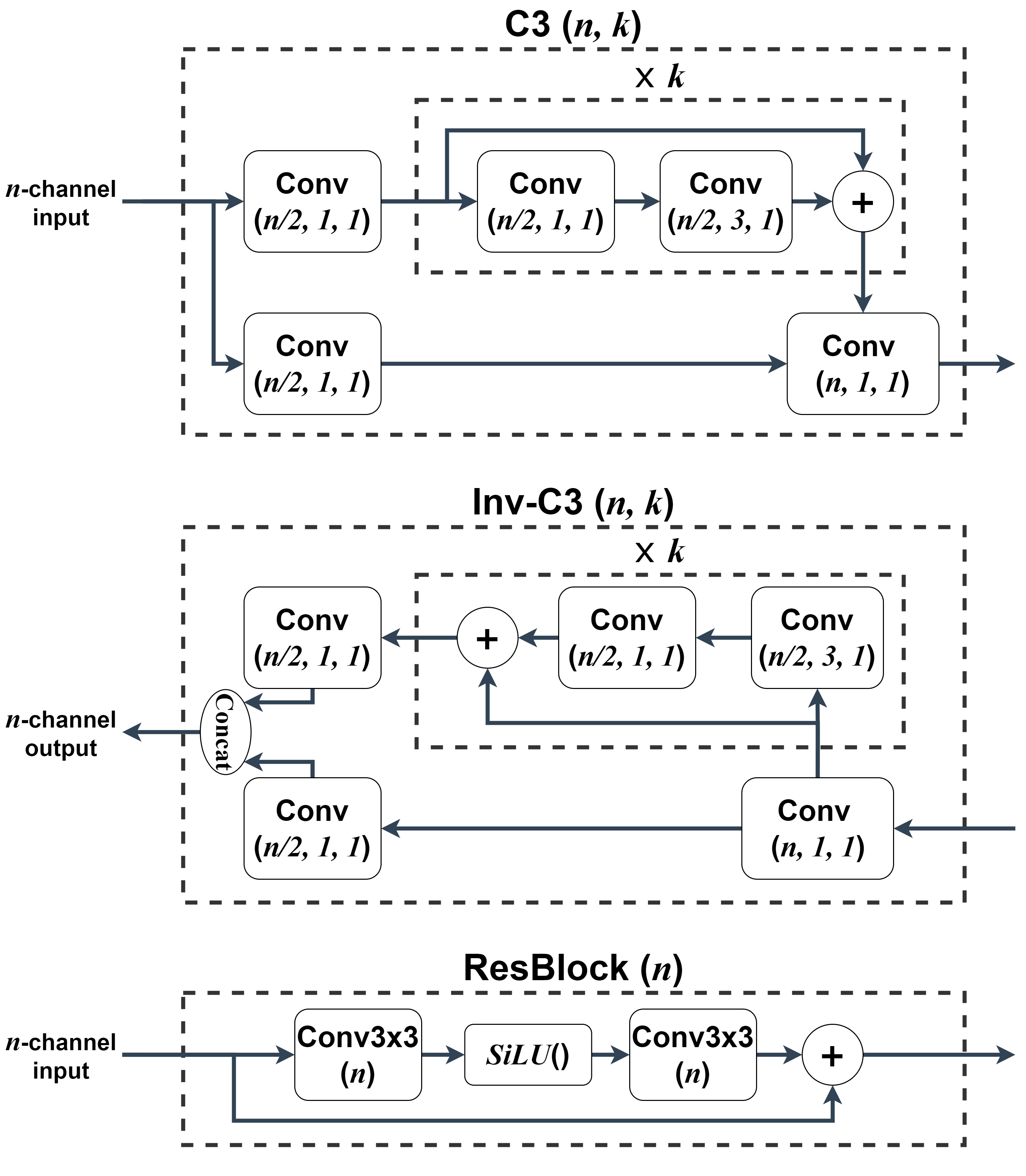}
    \caption{The architecture of some blocks shown in Fig.~\ref{fig:block_diagram}. $Conv(n,k,s)$ is the same as defined in the caption of Fig.~\ref{fig:block_diagram}.  $Conv3\times 3(n)$ is a single $3\times 3$ Convolution layer with stride=1 and $n$ output channels. ``$\times k$" indicates $k$ times repetition of the blocks specified by a dashed line.}
    \label{fig:sub_blocks}
\end{figure}

\subsection{Split Point}
To fit into the edge-cloud collaborative setting, the YOLOv5 model is split into two parts: the front-end, which is deployed on an edge device, and the back-end, which resides in the cloud. Choosing a split point is a design issue~\cite{kang2017neurosurgeon,eshratifar2019jointdnn}, which depends on energy considerations, computational resources at the edge, the type of connection between the edge and the cloud, and so on. Here, our selection of the split point is mainly based on information-theoretic considerations.

Let $\mathbf{X}$ be the input image and $\mathbfcal{Y}_i$ be the feature tensor at the $i$-th layer. 
Since $\mathbf{X} \to \mathbfcal{Y}_i \to \mathbfcal{Y}_{i+1}$ forms a Markov chain, by the data processing inequality~\cite{Cover_Thomas_2006} we have that
\begin{equation}
    I(\mathbf{X};\mathbfcal{Y}_i) \geq I(\mathbf{X};\mathbfcal{Y}_{i+1}),
    \label{eq:dpi}
\end{equation}
where $I(\cdot;\cdot)$ is the mutual information. Hence, deeper layers carry less information about the input and are therefore more resilient to model inversion attacks. Also, as shown in~\cite{scalable_coding_Hyomin}, deeper layers are more compressible. These arguments would suggest choosing a split point as deep as possible. 

On the other hand, limited computation and energy resources on the edge device favor selecting a shallower split point. Moreover, the YOLOv5m model branches out at layer 5, meaning that if we select the split point after layer 5, we would have to encode and transmit multiple feature tensors, which would increase both the complexity and the total bitrate. Hence, we choose to split the YOLOv5m model at layer 5. 

\subsection{Autoencoder}
At the split point, we insert an autoencoder, whose purpose is to both reduce the dimensionality and modify the features, such as to improve resistance to model inversion. This is a plug-and-play strategy and can be used for other models and tasks as well. The autoencoder is shown in Fig.~\ref{fig:block_diagram}: its encoder portion is referred to as AE and decoder as AD. They  consist of $Conv(n,k,s)$ layers, whose structure is specified in the caption of Fig.~\ref{fig:block_diagram}, and ResBlocks, whose structure is shown in Fig.~\ref{fig:sub_blocks}. The AE outputs the bottleneck feature tensor whose channel dimension ($64$) is lower than the input tensor's channel dimension ($192$), but the spatial dimensions are unchanged. This is done to preserve the spatial precision of subsequent object detection. The resulting bottleneck feature tensor is tiled, pre-quantized to 8-bits per element, and encoded using Versatile Video Coding (VVC)-Intra~\cite{VVC_overview}. At the cloud side, the encoded bitstream is decoded by a VVC decoder, then AD, then fed to the YOLOv5m back-end.

\subsection{Adversarial Training}
\label{subsec:adv_train}
Since the goal is to create bottleneck features with improved resilience towards model inversion, we construct an auxiliary \ac{DNN} model, called Reconstruction Network (RecNet), whose goal is to reconstruct the input image from bottleneck features.  As depicted in Fig.~\ref{fig:block_diagram}, the architecture of RecNet is roughly a mirror of the YOLOv5m front-end and AE. We train the autoencoder and  RecNet together in an adversarial manner~\cite{GAN}. Over the training process, RecNet tries to recover the input image from the bottleneck features as best it can, while the AE simultaneously tries to disrupt RecNet's performance by manipulating the generated bottleneck features. At the same time, both AE and AD attempt to keep the object detection accuracy high. Note that the pre-trained YOLOv5m model is kept intact, and its weights are frozen during the entire training process.
The training process for each batch of data is summarized in the following steps:
\begin{enumerate}
    \item The input $\mathbf{X}$ goes through the front-end, AE, and RecNet and the reconstruction loss is computed as follows:
     \begin{equation}
        \begin{split}
             L_{rec} = \frac{1}{n}\|\mathbf{X}-\widehat{\mathbf{X}}\|_1 & +  \frac{\beta}{n}\|S_x*\mathbf{X}-S_x*\widehat{\mathbf{X}}\|_1 \\ 
             & + \frac{\beta}{n}\|S_y*\mathbf{X}-S_y*\widehat{\mathbf{X}}\|_1,
        \end{split}
        \label{eq:L_rec}
     \end{equation}
     where $\widehat{\mathbf{X}}$ is the reconstructed input, $\|\cdot\|_1$ is the $\ell_1$-norm, $S_x$ and $S_y$ are the horizontal and vertical Sobel filters, respectively, $*$ is the convolution operator, and $n$ is the total number of tensor's elements in the batch. The value of $\beta$ is empirically set to $5$. We adopted the Sobel filter in the reconstruction loss in order to emphasize the edges since private information is usually associated with fine detail. As a result, RecNet becomes more powerful in reconstructing the edges, which will force AE to remove edge information from the bottleneck features.
     \item Gradient of $L_{rec}$ backpropagates only through the RecNet and updates its weights. Note that the autoencoder's weights are frozen at this step.
     \item The same batch of images goes through the whole network, and the total loss is computed as follows:
     \begin{equation}
             L_{tot} = L_{obj} - w\cdot L_{rec},
             \label{eq:L_tot}
     \end{equation}
     where $L_{obj}$ is the YOLOv5 object detection loss~\cite{glenn_jocher_2022_6222936}, and $w$ is the balancing weight between the reconstruction and object detection. We  empirically set it to $0.1$. 
     \item Gradient of $L_{tot}$ backpropagates only through the autoencoder and updates its weights. RecNet's weights are frozen in this step. Note that the negative sign of the $L_{rec}$ in~(\ref{eq:L_tot}) leads  AE to make reconstruction more difficult for RecNet. Meanwhile, the positive sign of $L_{obj}$ causes AE and AD to improve object detection accuracy.  
\end{enumerate}

\section{Experimental Results}
\label{sec:exp_res}
We trained our networks in several steps on the COCO-2017 object detection dataset~\cite{COCO} using an NVIDIA Tesla V100-SXM2 GPU with 32GB memory. In the first step, only the autoencoder got trained with $L_{obj}$ for 50 epochs. Next, we trained only the RecNet with $L_{rec}$ for 20 epochs, with the autoencoder's weights frozen to those obtained in the first step. Finally, with the autoencoder and RecNet initialized to the previously obtained weights, the adversarial training was conducted as described in Section~\ref{subsec:adv_train} for 40 epochs. 
In all the steps, we used Stochastic Gradient Descent (SGD) optimizer with the initial learning rate equal to 0.01, changing by a cosine learning rate decay \cite{bag_of_tricks_2019_CVPR} over the training. The object detection accuracy for the proposed autoencoder after training and the original YOLOv5m model is given in Table~\ref{tab:accuracy}.

\begin{table}[t]
    \caption{Results without feature compression}
    \begin{center}
    \begin{tabular}{|c|c|c|}
    \hline
    \textbf{Detection accuracy on COCO-2017}     &  \textbf{mAP@.5}  &  \textbf{mAP@.5:.95}\\
    \hline\hline
    \textbf{Proposed autoencoder}  &  $61.3$  &  $42.8$\\
    \hline
    \textbf{Original YOLOv5m}  &  $62.4$  &  $43.7$\\  
    \hline
    \end{tabular}
    \label{tab:accuracy}
    \end{center}
\vspace{-0.3cm}
\end{table}

\subsection{Resistance to model inversion}
As mentioned before, RecNet is an auxiliary DNN model exploited in the adversarial training stage. So, RecNet is not part of the final pipeline. In a real situation, however, if an adversary can get hold of the edge device, they can try to train their own input reconstruction model using input-bottleneck pairs. In order to test our model against this attack, we train a new, randomly initialized RecNet on the bottleneck features generated by the autoencoder obtained in the adversarial training stage.  This RecNet is trained with a $\ell_1$-norm loss (the first term in~(\ref{eq:L_rec})) and  is called RecNet-bottleneck. 
For comparison, we have also trained another RecNet, whose first three layers are removed, on the original YOLOv5m latent space (without the AE). We call this model RecNet-latent.

\begin{figure}[t]
    \centering
    
    \begin{subfigure}{0.3\columnwidth}
        \centering
        \includegraphics[width=\columnwidth]{figures/original_text.jpg}
    \end{subfigure} 
    \hfill
    \begin{subfigure}{0.3\columnwidth}
        \centering
        \includegraphics[width=\columnwidth]{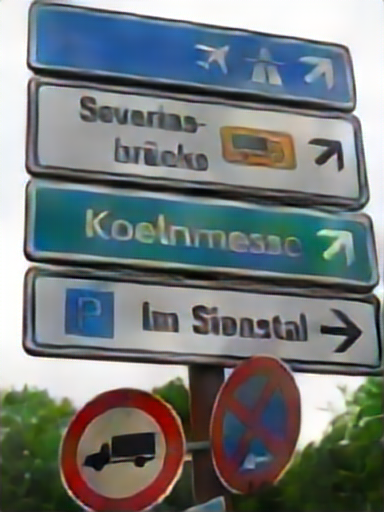}
    \end{subfigure} 
    \hfill
    \begin{subfigure}{0.3\columnwidth}
        \centering
        \includegraphics[width=\columnwidth]{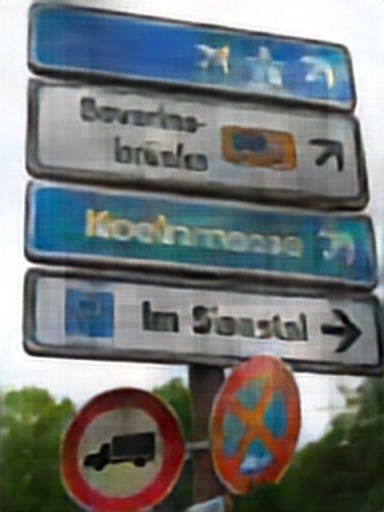}
    \end{subfigure}
    
    \par\smallskip
    
    \begin{subfigure}{0.3\columnwidth}
        \centering
        \includegraphics[width=\columnwidth]{figures/original_mom.jpg}
        \caption{}
    \end{subfigure}
    \hfill
    \begin{subfigure}{0.3\columnwidth}
        \centering
        \includegraphics[width=\columnwidth]{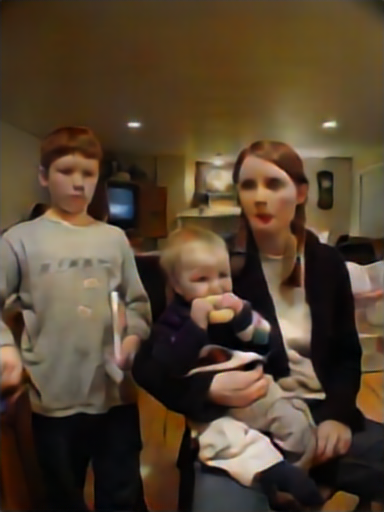}
        \caption{}
    \end{subfigure}
    \hfill
    \begin{subfigure}{0.3\columnwidth}
        \centering
        \includegraphics[width=\columnwidth]{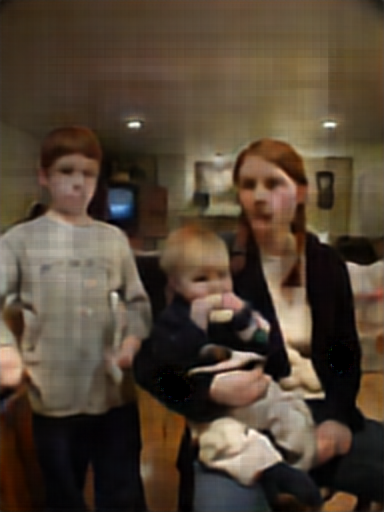}
        \caption{}
    \end{subfigure}
    
    
    \caption{Examples of  input reconstruction: (a) original image  (b) RecNet-latent's output (c) RecNet-bottleneck's output}
    \label{fig:vis-examples} 
    \vspace{-0.4cm}
\end{figure}

We test the resistance against model inversion attack by applying RecNet-latent to YOLOv5m features at layer 5 (without VVC compression) and applying RecNet-bottleneck to our bottleneck features (also without VVC compression). Input reconstruction performance is measured using conventional Peak Signal-to-Noise Ratio (PSNR), as well as a new quality metric called \textit{edge-PSNR} that emphasizes PSNR near the edges. To compute the edge-PSNR, the horizontal and vertical Sobel filters are applied to the original and reconstructed images to capture the image gradients in horizontal and vertical directions. Then, the magnitude of the gradient is considered as a new image based on which the edge-PSNR is calculated. 
The values of PSNR and edge-PSNR are given in Table~\ref{tab:quality}. These results show that our AE is able to remove some information required for input reconstruction, especially near the edges. In particular, reconstruction from our bottleneck features is 1.4 dB worse than reconstruction from YOLOv5m features, and this loss is mostly concentrated near the edges since edge-PSNR is 2.5~dB lower. 
Some visual examples are also provided in Fig.~\ref{fig:vis-examples}. As can be seen, the edges are more distorted in the output of RecNet-bottleneck, the text is not readable, and the faces and facial expressions are not easily recognizable.

\begin{table}[b]
    \caption{Input reconstruction results}
    \vspace{-0.2cm}
    \begin{center}
    \begin{tabular}{|c|c|c|c|}
    \hline
         &  \textbf{RecNet-latent}  &  \textbf{RecNet-bottleneck}  &  \textbf{Difference}\\
    \hline\hline
    \textbf{PSNR}  &  $21.78$~dB  &  $20.38$~dB  &  $-1.40$~dB \\
    \hline
    \textbf{edge-PSNR}  &  $25.61$~dB  &  $23.12$~dB  &  $-2.49$~dB\\  
    \hline
    \end{tabular}
    \label{tab:quality}
    \end{center}
\end{table}

\subsection{Feature compression results}
To measure the impact of feature compression, we encode the bottleneck features of the COCO validation images using VVC-Intra, specifically, its VVenC~\cite{VVenC} implementation with \textit{lowdelay-faster} preset. Prior to that, the $64$ channels in the bottleneck are clipped, 
quantized to $8$ bits, and tiled into an $8\times8$ matrix to create a gray-scale image. On the decoder side, the bitstream is decoded, 
and the resulting tensors are passed to AD and the YOLOv5m back-end for inference.

We found that most of the feature values in the bottleneck lie in the range of $[-6,6]$. As noted in~\cite{ccb_ojcas_2021}, feature compression performance can be improved via clipping. To this end, we tested three clipping ranges: 
$[-6,6]$, $[-3,3]$, and $[-1.5,1.5]$. We computed Bj{\o}ntegaard-Delta values \cite{bjontegaard2001calculation} between their associated Rate-Accuracy curves (not shown due to space constraints), where object detection accuracy is measured via mean Average Precision (mAP) at the IoU threshold of 0.5. QP values $\{34, 36, 38, 40, 41, 42\}$ were used to obtain these results. Using the performance for the $[-1.5,1.5]$ clipping range as the anchor, BD-Rate calculated based on the MPEG-VCM reporting template~\cite{mpeg_vcm_report} shows that clipping ranges $[-3,3]$ and $[-6,6]$ reduce the bitrate by $8.1\%$ and $7.1\%$ on average, respectively. Hence, we selected  
$[-3,3]$ as the clipping range for further experiments.

Finally, we compare our proposed method against direct coding of YOLOv5m layer 5 features, which we call the ``Anchor.'' 
Similar to encoding the bottleneck, the $192$ channels in the YOLOv5m latent space are clipped, quantized to 8 bits, tiled into a $12\times16$ matrix, and coded using VVC-Intra.

\begin{figure}[t]
    \centering
    \begin{subfigure}{0.6\columnwidth}
        \centering
        \includegraphics[width=\columnwidth]{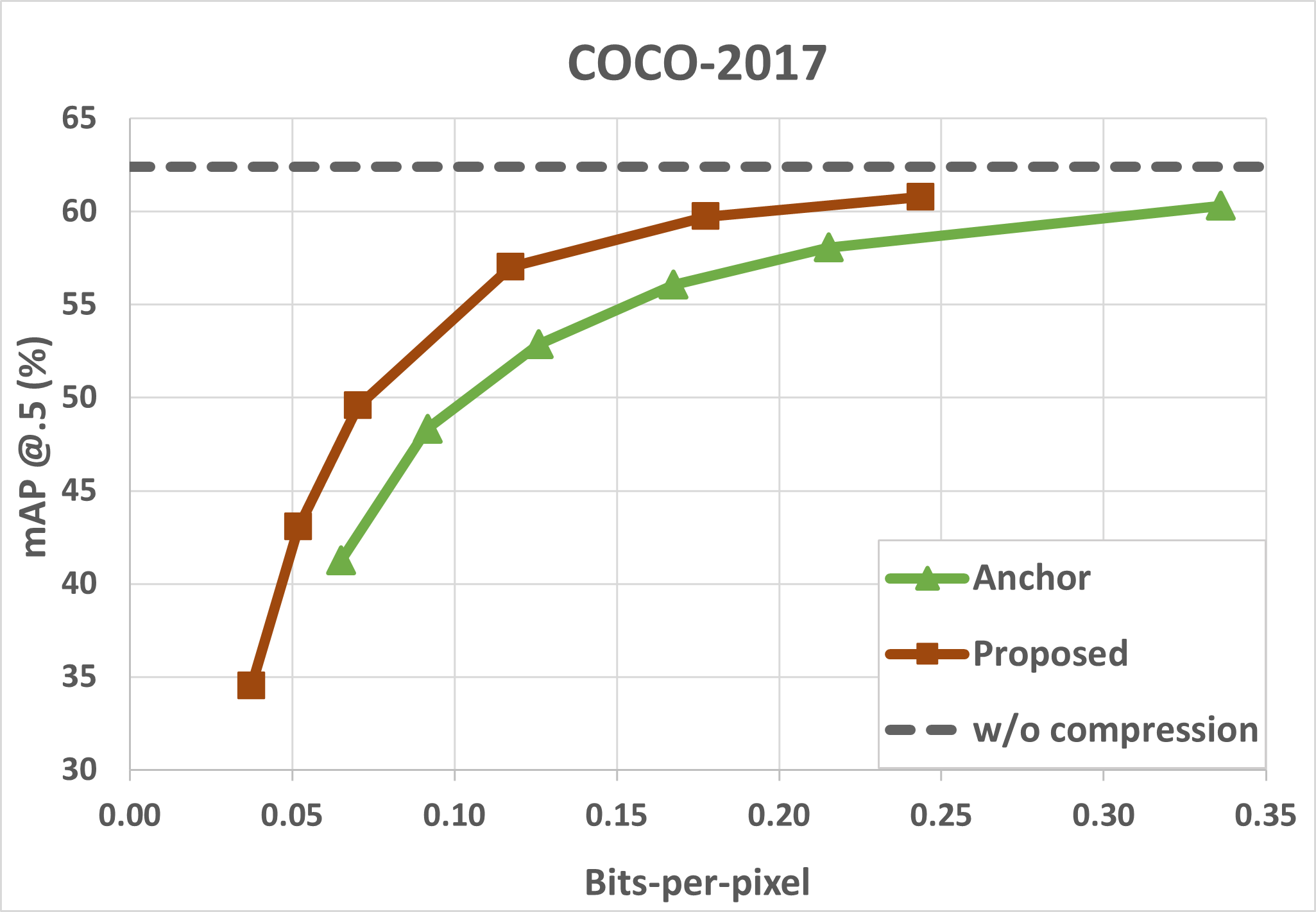}
        \caption{}
        \label{subfig:rate-acc}
    \end{subfigure} 
    \hfill
    \par\smallskip
    \begin{subfigure}{0.6\columnwidth}
        \centering
        \includegraphics[width=\columnwidth]{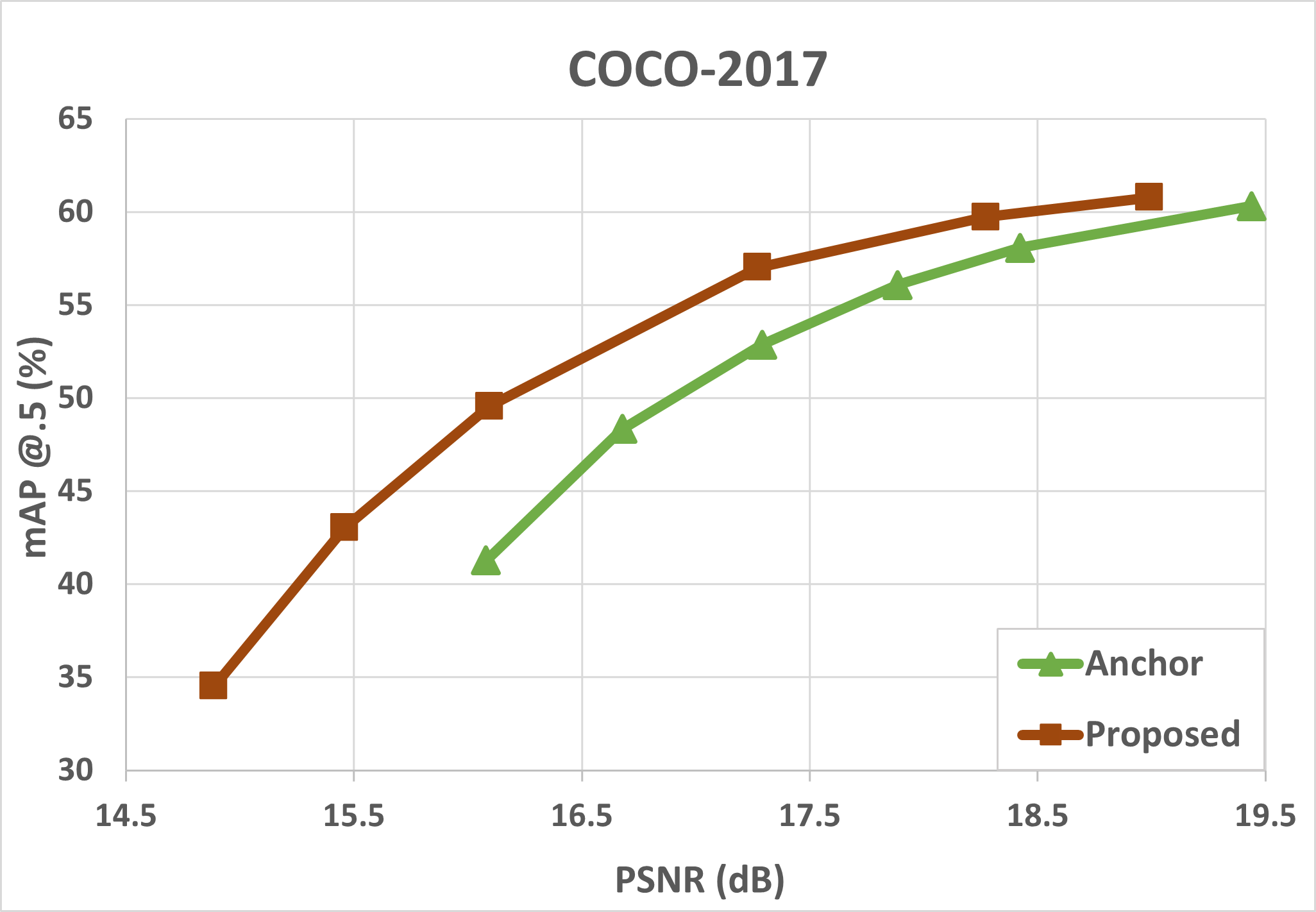}
        \caption{}
        \label{subfig:psnr-acc}
    \end{subfigure} 
    \caption{(a) Bits-per-pixel vs mAP@.5  (b) PSNR vs mAP@.5}
    \label{fig:charts}
    \vspace{-0.2cm}
\end{figure}

Fig.~\ref{fig:charts} shows the Rate-Accuracy and Quality-Accuracy curves obtained by QP $\in\{34, 36, 38, 40, 41, 42\}$ for the ``Proposed'' and QP $\in\{39, 41, 42, 43, 44, 45\}$ for the ``Anchor.'' 
BD-Rate between the curves presented in Fig.~\ref{subfig:rate-acc} is $-31.3\%$. Hence, our proposed method  reduces the bitrate by over $30\%$ on average for the equivalent accuracy. This reduction is not surprising given the reduced dimensionality of the bottleneck features. However, the novel benefit of the proposed method is the reduction of input reconstruction ability at the same accuracy, which is depicted in Fig.~\ref{subfig:psnr-acc}. Here, BD-PSNR between the curves is $-0.76$~dB, meaning that our bottleneck features lower the input reconstruction ability by about $0.8$~dB on average. As seen earlier, by design, most of the degradation is near the edges, which is important for privacy. 


\section{Conclusion}
\label{sec:conclusion}
In this paper, we presented a novel feature coding scheme for machines with improved resistance to model inversion attacks. The features of the YOLOv5m latent space were transformed into a lower-dimensional space using an autoencoder, which was trained in an adversarial manner to reduce the ability for input reconstruction. Visual and quantitative results showed that our method is able to degrade the quality of the recovered image using a model inversion attack, especially near the edges. Meanwhile, coding the features produced by our autoencoder resulted in more than $30\%$ bit savings, on average, at the same object detection accuracy compared to coding the original YOLOv5m features. 


\bibliographystyle{./IEEEtran}
\bibliography{./IEEEabrv, ./ms}

\end{document}